\documentclass[12pt]{article}

\usepackage{amssymb}

\newcommand{\C}{\mbox{\bf C}}

\newtheorem{definition}{{\bf Definition}}%[section]

\newtheorem{theorem}[definition]{{\bf Theorem}}
\newtheorem{proposition}[definition]{{\bf Proposition}}

\newtheorem{example}[definition]{{\bf Example}}

\newenvironment{proof}{\noindent {\bf Proof. }}{\hspace{\fill}$\Box$}

\newcommand{\mapright}[1]{\smash{\mathop{
 \hbox to 1cm{\rightarrowfill}}\limits^{#1}}}
\newcommand{\mapleft}[1]{\smash{\mathop{
 \hbox to 1cm{\leftarrowfill}}\limits^{#1}}}
\newcommand{\maprightd}[1]{\smash{\mathop{
 \hbox to 1cm{\rightarrowfill}}\limits_{#1}}}
\newcommand{\mapleftd}[1]{\smash{\mathop{
 \hbox to 1cm{\leftarrowfill}}\limits_{#1}}}

\title{Generalized Partitioned Quantum Cellular Automata and
Quantization of Classical CA}
\author{
Shuichi INOKUCHI\footnote{Faculty of Mathematics, Kyushu University, 
E-mail:inokuchi@math.kyushu-u.ac.jp} 
\and 
Yoshihiro MIZOGUCHI\footnote{Faculty of Mathematics, Kyushu
University, E-mail:ym@math.kyushu-u.ac.jp}}

\begin{document}

\maketitle
\begin{abstract}
 In this paper, in order to investigate natural transformations from
 discrete CA to QCA, we introduce a new formulation of finite cyclic QCA
 and generalized notion of partitioned QCA. According to the
 formulations, we demonstrate the condition of local transition
 functions, which induce a global transition of well-formed
 QCA. Following the results, extending a natural correspondence of
 classical cells and quantum cells to the correspondence of classical CA
 and QCA, we have the condition of classical CA such that CA generated
 by quantization of its cells is well-formed QCA. Finally we report
 some results of computer simulations of quantization of classical
 CA. 

\end{abstract}
\section{Introduction}
J. Watrous introduced the notion of quantum cellular automata(QCA)
and showed that any quantum Turing machine can be efficiently simulated
by a QCA with constant slowdown in 1995.

CA with quantum cells is well-formed QCA if and only if its global
transition function is unitary.
Generally quantization of cells of a classical CA dose not
make it become QCA, because usually classical CA dose not have reversibility.
Morita and Harao show that we can get reversible CA by partition a cell
into three part and partitioned CA(PCA) can simulate non-partitioned
CA(NPCA)\cite{morita89}.
But there is not a trivial inclusion relation between PCA and NPCA.

In this paper, we introduce a new formulation of finite cyclic QCA
and generalized notion of partitioned QCA in order to investigate 
natural transformations from discrete CA to QCA.
According to the formulations, we demonstrate a condition of a local
transition function, which induce a well-formed QCA.
A natural correspondence of classical cells and quantum cells can be
extended to the correspondence of classical CA and QCA.
If a classical CA satisfies our conditions then the extended QCA is
well-formed.
Finally we report some results of computer simulations of quantization of
classical CA.
\section{Preliminaries}

Let $Q$ be a set of states of cells and $|Q|=s$.
We consider $Q^{n}$ is the set of configurations of CA where $n$ is the size of CA.
$q_{i}$ denotes the $i$th element of the configuration $q\in Q^{n}$, and assume that $q_{0}=q_{n}$ and $q_{n+1}=q_{1}$.

Before considering computing process of quantum states, we recall that of deterministic states.
We use the set of all subset of $Q$, that is, the set $2^Q$ of all functions from  $Q$ to $2=\{0,1\}$ to represent nondeterministic states.

A element $q$ of $Q$ is normally considered as an element $\{q\}$ of $2^Q$.
Let $\Sigma$ be the finite set of input characters, then
a transition function of input characters of deterministic finite automata 
is provided as $\delta:Q \times \Sigma \to Q$, and expanded naturally into 
the transition function $\delta_*: Q \times \Sigma^* \to Q$ of input strings.
Let $[\delta]:Q \times \Sigma \to 2^Q$ be a function defined by $[\delta](q,\alpha)=[\delta(q,\alpha)]$.

The function $[\delta]$ is a state transition function of
nondeterministic finite automata and we can expand $[\delta]$ into
$[\delta]_*: Q \times \Sigma^* \to 2^Q$.
Recall $[\delta_*]:Q \times \Sigma^* \to 2^Q$ be a function defined by
$[\delta_*](q,\alpha)=[\delta_*(q,\alpha)]$, and
we can show that $[\delta_*] = [\delta]_*$ easily.
This shows that the set of all deterministic finite automata are
included in the set of all nondeterministic finite automata naturally.

We can consider quantum states as generalized states of classical state
and extend to a quantum formulation of computer system. But a classical
computer system is not always a quantum computer system generally,
because a quantum computing process should be a unitary operator and
every classical computing process is not so.

A quantum state denotes a function from a finite set $Q$ to a set of
complex numbers $\C$ and $\C^Q$ is denoted by the set of all functions
from $Q$ to $\C$. For $g\in Q$ we define $[q] \in \C^Q$ as follows;
$$
[q](x) = \left\{
\begin{array}{ll}
1 &  (x=q) \\
0 &  (x \not= q)
\end{array}
\right.
$$
$\C^Q$ is a linear space on $\C$ such that its bases is $Q$. An inner product in $\C^Q$ is defined by 
$\displaystyle{\langle p,q \rangle = \sum_{x \in Q} (p(x)\cdot q(x))}$
where $p,q\in \C^Q$. 

A linear space $\C^{(Q^n)}$ on $\C$ is considered as a tensor product $\displaystyle{\underbrace{\C^Q \otimes \C^Q \otimes \cdots \otimes \C^Q}_n}$, that is,
for $p \in Q^n$ $[p] = [p_1] \otimes [p_2] \otimes \cdots \otimes [p_n]$.
For $p, q \in Q^n$ the inner product $\langle [p],[q]\rangle $ on
$\C^{(Q^n)}$ is as follows;
\begin{eqnarray*}
 &   & \langle [p],[q] \rangle\\
 & = & \sum_{x \in Q^n}([p](x)\cdot [q](x)) \\
 & = & \sum_{(x_1,x_2,\ldots,x_n) \in Q^n}
	(([p_1](x_1) [p_2](x_2) \cdots [p_n](x_n)) \\
 &   & \hspace{2cm} \cdot
	([q_1](x_1) [q_2](x_2) \cdots [q_n](x_n))) \\
 & = & \sum_{x_1 \in Q} ([p_1](x_1) [q_2](x_1)) \cdot
	\sum_{x_2 \in Q} ([p_2](x_2) [q_2](x_2)) \\
 &   & \hspace{1cm} \cdots
	\sum_{x_n \in Q} ([p_n](x_n) [q_n](x_n)) \\
 & = & \langle [p_1],[q_1] \rangle \langle [p_2],[q_2] \rangle \cdots
	\langle [p_n],[q_n] \rangle.\\
\end{eqnarray*}
For a function $F:Q^n \to \C^{(Q^n)}$, we define functions
$\alpha_F:Q^n\times Q^n \to \C$ and $\overline{F}:\C^{(Q^n)} \to
\C^{(Q^n)}$
by $\alpha_F(p,q)=F(p)(q)$
and $\displaystyle{\overline{F}(X)= \sum_{q \in Q^n}(X(q)(F(q)))}$.
We call $\overline{F}$ is {\it unitary} if $||\overline{F}(X)||=1$
for any $X \in \C^{(Q^n)}$ such that $||X||=1$.

Since $Q$ is finite, we can label elements of $Q$ numbers from 1 to $s$
and also elements of $Q^n$ numbers from 1 to $s^n$ by lexicographical ordering.
We define a $s^n \times s^n$ matrix $(\alpha_{i j})$ by 
$\alpha_{i j}=\alpha_F(p,q)$ where numbers of elements $p$ and $q$ are
$i$and $j$.

\begin{proposition}\label{prop:unitaly}
If the matrix $(\alpha_{i,j})$ is unitary then $\overline{F}$ is unitary.
\end{proposition}
\begin{proof}
Assume that $(\alpha_{i j})$ is a unitary matrix and
$\langle X,X \rangle = 1$.
\begin{eqnarray*}
& & \langle \overline{ F}(X), \overline{ F}(X) \rangle\\
 & = & \langle \sum_{p \in Q^n}X(p)\overline{ F}(p), 
            \sum_{q \in Q^n}X(q)\overline{ F}(q) \rangle \\
& = & 
   \sum_{p \in Q^n} \sum_{q \in Q^n} \langle X(p)\overline{ F}(p),
           X(q)\overline{ F}(q) \rangle \\
& = & \sum_{p \in Q^n} (X(p) \sum_{q \in Q^n} (X(q)
           \sum_{r \in Q^n} (\overline{ F}(p)(r) 
                  \cdot \overline{ F}(q)(r)) ))\\
& = & \sum_{p \in Q^n} (X(p) \sum_{q \in Q^n} (X(q)
           \sum_{r \in Q^n} (\alpha(p,r) \cdot \alpha(q,r)) ))\\
& = & \sum_{p \in Q^n} (X(p) \sum_{q \in Q^n} (X(q)
   \sum_{r \in Q^n} (\alpha(p,r) \cdot \overline{\alpha(r,q)}))) \\
& = & \sum_{p \in Q^n} (X(p) \cdot X(p)) \\
& = & 1
\end{eqnarray*}
So $\overline{F}$ is unitary.
\end{proof}

\begin{proposition}
Let $\hat{\sigma}:\C^{Q^n} \to \C^{Q^n}$ be defined by 
$\hat{\sigma}(X)(p)=X(\sigma(p))$ where $\sigma:Q^n \to Q^n$.
Then the followings are equivalent.
  \begin{itemize}
    \item $\hat{\sigma}=\overline{[\sigma]}$ and $\hat{\sigma}$ are unitary.
    \item $\sigma$ is a bijection.
  \end{itemize}
\end{proposition}

A classical CA is a transition system in $Q$ defined by a global
transition function $F:Q^n \to Q^n$ where 
$F(q)_i=f(q_{i-1},q_i,q_{i+1})$ and $f:Q \times Q \times Q \to Q$ is a
local transition function.

When $Q=\{0,1\}$,
a local transition function is defined by the eight values
$f(0,0,0)=r_0$, $f(0,0,1)=r_1$, $f(0,1,0)=r_2$,
$f(0,1,1)=r_3$, $f(1,0,0)=r_4$, $f(1,0,1)=r_5$,
$f(1,1,0)=r_6$ and $f(1,1,1)=r_7$ ($r_i=0,1$).
The rule number $R$ of a local transition function $f$ is defined by 
$$R=2^7 r_7\!+\! 2^6 r_6\!+\! 2^5 r_5\!+\! 2^4 r_4\!+\! 2^3 r_3\!+\! 2^2
r_2\!+\! 2^1 r_1\!+\! r_0.$$
The local transition function of rule number $R$ is denoted by $f_R$.
The local transition rules with rule number 204,240 and 170 
are illustrated as follows;
{\footnotesize
\begin{center}
\begin{tabular}{|l|cccccccc|} \hline
     & 111 & 110 & 101 & 100 & 011 & 010 & 001 & 000 \\
204  &  1  &  1  &  0  &  0  &  1  &  1  &  0  &  0  \\ \hline
     & 111 & 110 & 101 & 100 & 011 & 010 & 001 & 000 \\
240  &  1  &  1  &  1  &  1  &  0  &  0  &  0  &  0  \\ \hline
     & 111 & 110 & 101 & 100 & 011 & 010 & 001 & 000 \\
170  &  1  &  0  &  1  &  0  &  1  &  0  &  1  &  0  \\ \hline
\end{tabular}
\end{center}
}
$f_{204}$, $f_{240}$ and $f_{170}$ are identity, shift-right and shift-left functions respectively.

%%%%%%%%%%%%%%%%%%%%%%%%%%%%%%%%%%%%%%%%%%%%%%%%%%%%%%%%%%%%
\section{Quantum Cellular Automata}

The global transition function $F:Q^n \to \C^{(Q^n)}$ is defined by 
$F(q)(x)=f(q_0,q_1,q_2)x_1 \cdot
         f(q_1,q_2,q_3)x_2 \cdots
         f(q_{n-1},q_n,q_{n+1})x_n$.
For a local transition function $f:Q \times Q \times Q \to \C^Q$,
A transition system in $\C^{(Q^n)}$ is defined by
$\overline{F}:\C^{(Q^n)} \to \C^{(Q^n)}$.
The local transition function $f$ is called 'forming a quantum cellular
automaton' if $\overline{F}$ is unitary.

Let $F_f:Q^n \to Q^n$ be a transition function of a local transition
function $f:Q^3 \to Q$, and$[F_f]:Q^n\to \C^{Q^n}$ a function such that
$[F_f](x)=F_f(x)$ for $x\in Q^n$.
And we let $[f]:Q^3 \to \C^Q$ be a local transition function of QCA such
that $[f](x)=f(x)$ for $x\in Q$, and $F_{[f]}:Q^n \to \C^{(Q^n)}$ its
global transition function.
Then we can show that $[F_f] = F_{[f]}$ by easy computation.
But a local transition functions $f:Q^3 \to Q$ does not always form a
QCA, that is, a $\overline{[F_f]}$ is not always unitary.

\begin{proposition}
$[f]$ is forming a quantum cellular automata if and only if $F_f:Q^n \to Q^n$ is a bijection.
\end{proposition}

%%%%%%%%%%%%%%%%%%%%%%%%%%%%%%%%%%%%%%%%%%%%%%%%%%%%%%%%%%%%
\section{Partitioned Quantum Cellular Automata}

We define functions $G:Q^n \to \C^{(Q^n)}$
and $\lambda:Q \times Q \to \C$ by $G(q)(x)=g(q_1)x_1 \cdot g(q_2)x_2
\cdots g(q_n)x_n$ and $\lambda(p,q)=g(p)(q)$ for a function $g:Q \to C^{Q}$.

We label the elements of $Q^n$ numbers from 1 to $s^n$ and 
define a $s^n \times s^n$ matrix $(\alpha_{i j})$ from $\alpha_G:Q^n
\times Q^n \to \C$. And we label the elements of $Q$ numbers from 1 to
$s$ and define a $s \times s$ matrix $(\lambda_{i j})$ from the function $\lambda$.

\begin{proposition}
$(\lambda_{i j})$ is a unitary matrix if and only if $(\alpha_{i j})$ is a unitary matrix.
\end{proposition}

\begin{proposition}
If $\sigma:Q^n \to Q^n$ is a bijection, then the followings hold;
\begin{description}
\item[(i)] $\overline{G \circ \sigma} = 
       \overline{G} \circ \hat{\sigma}$.
\item[(ii)] $\overline{G \circ \sigma}$ is unitary if and only if
       $\overline{G}$ is unitary.
\end{description}
\end{proposition}

$$
\begin{array}{ccccc}
Q^n & \mapright{\sigma} & Q^n & \mapright{G} & \C^{(Q^n)} \\
{\Big\downarrow} && {\Big\downarrow} && {\Big\Updownarrow} \\
\C^{(Q^n)} & \maprightd{\hat{\sigma}} & \C^{(Q^n)} 
                           &\maprightd{\overline{G}}& \C^{(Q^n)} \\
\end{array}
$$
\begin{theorem}\label{th1}
The composition function $f=g \circ e: Q^3 \to \C^Q$ of functions
$e:Q^3 \to Q$ and $g:Q \to \C^Q$ is forming a quantum cellular automaton
if both of the following two conditions hold:
\begin{description}
\item[(i)] $F_e:Q^n \to Q^n$ is a bijection.
\item[(ii)] The matrix $(\lambda_{i j})$ defined from
$g:Q \to \C^Q$ is unitary.
\end{description}
\end{theorem}

\begin{example}\label{exam:partition}
 Let $Q=L\times M \times R$ for finite sets $L$, $M$ and $R$.
 We define $e:Q^3 \to Q$ and $g:Q \to \C^Q$ by
 $e(((l_1,m_1,r_1),(l_2,m_2,r_2),(l_3,m_3,r_3)))=(l_3,m_2,r_1)$ and
 $g(q)=[q]$.
 Then the composition function $f=g \circ e$ is forming a quantum
 cellular automaton. Because $F_e:Q^n \to Q^n$ is a bijection, 
 $F_e(((l_i,m_i,r_i)))_j = (l_{j+1},m_j,r_{j-1})$, and 
 $(\lambda_{i j})$ defined from $g$ is an identity matrix.
\end{example}

In example \ref{exam:partition}, we can replace $g$ to another function
$g:Q \to \C^Q$ where the matrix $(\lambda_{i j})$ defined from $g$ is
unitary. On that occasion $f:(L \times M \times R)^3 \to (L \times M
\times R)$ is also forming a quantum cellular automaton. Consequently a
partitioned quantum cellular automaton introduced in \cite{watrous95} is
demonstrated as a special case of our general formulation.

\begin{example}\label{ex1}
Let $Q=\{0,1\}$, 
$e:Q^3 \to Q$ be a function such that $F_e:Q^n \to Q^n$ is a bijection, and 
$\Lambda=(\lambda_{i j})$ defined from $g:Q \to \C^Q$ be as follows;
$$
\Lambda=
 \left( 
 \begin{array}{cc}
	\cos \theta & - \sin \theta \\
	\sin \theta & \cos \theta 
 \end{array}
 \right). 
$$
Then the local transition function defined from $f=g \circ e$ forms a
quantum cellular automaton.
This shows a quantum cellular automaton formed by a synthesised function
of a local transition function $e$ and its reverse function.
That is, if $\theta=0$ then $f=e$ and if $\theta=\frac{\pi}{2}$ then $f$
is the reversed function of $e$, so we can consider $f$ for 
$0<\theta<\frac{\pi}{2}$ as a synthesised function of two classical
local functions.
\end{example}

\begin{example}
 Let $Q=\{0,1\} \times \{0,1\}$,\\
 $e((a_1,b_1),(a_2,b_2),(a_3,b_3))=(a_1,b_3)$, and 
 $\Lambda=(\lambda_{i j})$ defined by $g:Q \to \C^Q$ be as follows;
 $$\Lambda = 
 \left( 
 \begin{array}{cccc}
	1 & 0 & 0 & 0 \\
	0 & 1 & 0 & 0 \\
	0 & 0 & 0 & 1 \\
	0 & 0 & 1 & 0 
 \end{array}
 \right).
 $$
 Then $f=g \circ e$ forms a QCA and
 $f((a_1,b_1),(a_2,b_2),(a_3,b_3))=(a_1,a_1 \oplus b_3)$.
\end{example}

\section{Computer Analysis}
The following is a table of the size $n$ of CA and rule number $R$ where the local transition function $f_R:Q^{3}\to Q$ forms QCA ($Q=\{0,1\}$, $3\leq n\leq 22$ and $128\leq R\leq 255$).
We note that if $f_R$ forms QCA then $f_{255-R}$ also forms QCA.
{\footnotesize
\begin{center}
\begin{tabular}{|l|l|} \hline
Size $n$ & Rule number $R$ \\ \hline \hline
3 & 142, 154, 156, 166, 170, 172, 178, 180, 184 \\
  & 198, 202, 204, 210, 212, 216, 226, 228, 240 \\ \hline
4, 8, 10, 14  & 150, 170, 204, 240 \\
16, 20, 22 &               \\ \hline
5, 7, 11, 13, 19 & 150, 154, 166, 170, 180, 204, 210, 240 \\ \hline
9, 15, 21 & 154, 166, 170, 180, 204, 210, 240 \\ \hline
6, 12, 18 & 170, 204, 240 \\ \hline
\end{tabular}
\end{center}
}
In the case of size 6, 12 and 18, trivial transitions, that is, identity, right shift and left shift functions only form QCA. In other case, there is a nontrivial transition function forming QCA.

From the above table, we can guess the following table of sizes of CA and rule numbers of local transition functions forming QCA, but we have not proved it yet.
{\footnotesize
\begin{center}
\begin{tabular}{|l|l|} \hline
Size ($k \ge 1$) & Rule number \\ \hline \hline
$6k \pm 2$  & 150, 170, 204, 240 \\ \hline
$6k \pm 1$ & 150, 154, 166, 170, 180, 204, 210, 240 \\ \hline
$6k + 3$ & 154, 166, 170, 180, 204, 210, 240 \\ \hline
$6k$ & 170, 204, 240 \\ \hline
\end{tabular}
\end{center}
}
\begin{example}\label{ex2}
When the size of CA is $4$ or $5$, 
the local transition function $f_{150}(x,y,z)=x+y+z (\mbox{\bf mod} 2)$
forms a quantum CA.\\
Let $F_f$ be the global transition function, then the following hold;
\begin{eqnarray*}
  F_{f}(x)_{i} & = & x_{i-1}+x_{i}+x_{i+1}\\
  F_{f}^{2}(x)_{i} & = & x_{i-2}+x_{i-1}+x_{i} \\
   && +x_{i-1}+x_{i}+x_{i+1}\\
   && +x_{i}+x_{i+1}+x_{i+2}\\
  & = & x_{i-2}+x_{i}+x_{i+2}\\
  F_{f}^{3}(x)_{i} & = &x_{i-3}+x_{i-2}+x_{i-1}\\
  &&+x_{i-1}+x_{i}+x_{i+1} \\
  &&+x_{i+1}+x_{i+2}+x_{i+3}\\
  & = & x_{i-3}+x_{i-2}+x_{i}+x_{i+2}+x_{i+3}
\end{eqnarray*}

In the case that the size of CA is $4$,
$F_{f}^{2}(x)_{i}=x_{i+2}+x_{i}+x_{i+2}=x_i$.
So $F_{f}^{2}(x)=x$.
In the case that the size of CA is $5$,
$F_{f}^{3}(x)_{i}=x_{i+2}+x_{i+3}+x_i+x_{i+2}+x_{i+3}=x_i$.
So $F_{f}^{3}(x)=x$.
Namely there exists $y$ such that $F_{f}(y)=x$ for any $x \in Q^n$ $(n=4,5)$,
and $F_f$ is a bijection. So $f$ forms a QCA.

\end{example}

\section{Related Works and Conclusion}
Cellular automata dealt in this paper is finite cyclic CA and different
from CA without boundary, that is, infinite CA dealt by
Watrous\cite{watrous95}, Morita and Harao\cite{morita89}. Because the
size of CA is finite it does not have the universal computability\cite{Toff77,birth92,deut85}.
But the conditions of local transition functions forming QCA is
formulated clearly in our framework.

Injectivity of global maps of classical CA is an essential property
for extending to a QCA.
The injectivity are considered in \cite{ito83,maruoka76,maruoka79,Toff77}
for classical CA without boundary and in \cite{harao72,dow97,ichikawa82}
for classical finite cyclic CA.

A further direction of this study will be to
consider properties on construction and synthesis of general quantum
computer system by examining construction and synthesis of local
transition function of realizable QCA.

\end{document}